# Vortex Mechanics in Thin Elliptical Ferromagnetic Nanodisks

C. E. Zaspel

Department of Environmental Sciences

University of Montana-Western

Dillon, MT 59725, USA

The magnetostatic energy is calculated for a magnetic vortex in a noncircular elliptical nanodisk. It is well-known that the energy of a vortex in the circular disk is minimized though an ansatz that eliminates the magnetostatic charge at the disk edge. Beginning with this ansatz for the circular disk, a conformal mapping of a circle interior onto the interior of an ellipse results in the magnetization of the elliptical disk. This magnetization in the interior of an ellipse also has no magnetostatic charge at the disk edge also minimizing the magnetostatic energy. As expected the energy has a quadratic dependence on the displacement of the vortex core from the ellipse center, but reflecting the lower symmetry of the ellipse. Through numerical integration of the magnetostatic integral a general expression for the energy is obtained for ellipticity values from 1.0 to about 0.3. Finally a general expression for the gyrotropic frequency as described by the Thiele equation is obtained.

Email: craig.zaspel@umwestern.edu

Tel. 406-683-7366

Fax 406-683-7493

**I. Introduction**

In submicron ferromagnetic insulators the combined effects of the exchange interaction and the magnetostatic dipole-dipole interaction tends to result in curling of the magnetization to minimize the magnetostatic surface charge at the nanocontact edge. This can result in a vortex structure where the magnetization is in-plane over most of the nanoparticle. There is also a small out-of- plane vortex core eliminating the exchange singularity at the vortex center. Vortex structures can have applications in the areas of information storage as well as microwave frequency signal generation and processing. Owing to these potential applications, magnetic vortex structures have received a large amount of experimental and theoretical study during the past ten to fifteen years. One of the most prominent dynamical effects is gyrotropic precession of the vortex core, which has been experimentally [1-4] observed in thin circular disks. In these systems the vortex core can be forced off-center by an in-plane magnetic field pulse and gyrotropic precession is imaged as spiral motion of the core as it returns to the disk center as a result of dissipation. Outside of advancing the theoretical understanding of nanoscale magnetization dynamics, recent applied applications include vortex-based nanoscale oscillators driven by spin-torque [5,6] as well as synchronization [7,8] of coupled spin-torque oscillators. The theoretical description of these systems requires knowledge of the magnetostatic (MS) energy from which the gyrotropic precession frequency can be derived. For the special case of the circular disk the form of the MS energy [7, 8] and gyrotropic frequency [1] are well known to depend only on the disk radius and thickness, but noncircular elliptical oscillators are also widely used [5, 9-11]. In particular, experimental results from Refs. [10,11] show how the gyrotropic frequency depends on the ellipse thickness as well as the ellipse axes; moreover, these data were also close to predictions by micromagnetic calculations. Since micromagnetic calculations are restricted to a special case, it can be very useful to have an expression giving the gyrotropic frequency for general values of the ellipse parameters, which is the focus of this article.

There has been much successful theoretical work to explain gyrotropic motion in the circular disk based on the Thiele equation [11, 12] which is obtained from the Landau-Lifshitz equation using collective coordinates. In this case the vortex behaves like a particle with its position in the disk given by the vortex core position. The Thiele equation typically includes the gyrotropic force along with conservative forces derivable from a potential as well as the nonconservative forces from dissipation. From these forces particle-like vortex dynamics can be obtained [1,14-17] to get the gyrotropic frequency as well as the radius of the vortex core orbit in the nanostructure. Calculation of the conservative force from the MS potential energy has been the major and most difficult part of theoretical descriptions of gyrotropic motion. Because of this difficulty analytical calculations can be done only for the highest symmetry case of the thin circular disk, but even for this highest symmetry case magnetostatic integrals must be approximated or evaluated numerically. For both analytical and numerical methods the MS energy is seen to depend on displacement, $X$ of the vortex core from the disk center having the general form, $W_{MS} \sim L^2/R \, X^2$ where $L$ is the disk thickness and $R$ is the disk radius. Typically this $L^2/R$ dependence of the MS energy results in a gyrotropic frequency of the form $\omega_{GT} \sim L/R$, which is also observed [1-4] experimentally. To find the quadratic dependence of the MS energy on $X$ usually a pole-free model [1] is assumed where an image vortex results in a magnetization distribution without a component normal to the disk edge. This magnetization distribution has no magnetic edge charge which tends to minimize the MS energy, but the volume MS charge is increased as the vortex core moves off of the disk center. Moreover, the nonzero exchange energy is also minimized through the vortex ansatz. Using this simple model there has been very good agreement between measured and predicted gyrotropic frequencies for various disk thicknesses and radii.

For ferromagnetic noncircular elliptical nanoparticles the dynamics of single [6] and multiple vortices [9-11] have been experimentally observed including effects such as gyrotropic motion, however, theoretical analysis has been lacking owing to increased analytic and numerical complexity owing to the lower symmetry. Recently simulations [18] have been done to determine the form of the MS energy as a function of the ellipticity, but the nature of this work is applicable to specific ellipses and cannot be applied to arbitrary sizes and ellipticity. Therefore, it would be useful to have a simple expression relating the MS energy to quantities such as the ellipse thickness along with the semi-major and minor axes. In the following the energy of a vortex confined to an ellipse is obtained with the edge magnetization parallel to the ellipse edge thereby minimizing the MS energy. This is accomplished through a conformal mapping of the interior of a circular disk to the interior of the ellipse so that the energy can be calculated through numerical evaluation of the MS integral. This article is arranged as follows: First the magnetization in the ellipse is obtained through the conformal mapping of the circle to the ellipse with the appropriate boundary conditions at the ellipse edge. Second, this magnetization is used to obtain the MS energy as a function of the vortex displacement from the ellipse center for small displacements of the vortex core for different ellipse parameters. Third, using these numerical data a general form of the MS energy is determined. Finally the gyrotropic frequency is obtained for vortices in elliptical nanoparticles in general.

**II. Mapping from circle to ellipse**

We begin with the circular cylinder that is sufficiently thin ($L \ll R$) so that the magnetization can be assumed to be uniform throughout its thickness. The magnetization is expressed in term of a normalized magnetization, $\vec{m}$ with $\vec{M} = M_s(m_x, m_y, m_z)$ and $M_s$ is the saturation magnetization. The component $m_z$ is perpendicular to the disk plane which will be nonzero at the vortex core. Moreover, the core is typically the size of the exchange length (5 nm for permalloy) so compared to a typical nanoparticle size (few hundred nm) the small core can be approximated as a δ-function and its contribution to the MS energy can be neglected resulting in the condition $m_x^2 + m_y^2 = 1$ over the vortex. As a result of the small core relative to the other dimensions, the exchange energy is approximately constant and does not significantly contribute to the calculation of the gyrotropic frequency.

Let us first give the magnetization in the circular disk which will be transformed to the ellipse. In this case the magnetization can be expressed as a function of the complex variable $t = t_1 + it_2$ with the components of the magnetization given by

$$m_x + im_y = \frac{2w}{1+w\bar{w}}, \tag{1}$$

where the bar indicates complex conjugation and $w$ is an analytic function. It is remarked that any analytic function will minimize the exchange energy so the next step is to find a less general function that will minimize the MS energy subject to the in-plane constraint of the magnetization. For the vortex magnetization distribution the function $w$ is given by

$$w = \frac{f(t)}{\sqrt{f(t)\bar{f}(\bar{t})}}, \tag{2}$$

which gives the meron type structure. The form in Eq. (2) does not include the small vortex core and for the reason stated above, the core can be neglected in the MS energy calculation. The analytical function, $f(t)$ describing the vortex with no edge magnetostatic charge [19] is the complex function

$$f(t) = i\left(t - (\alpha + \bar{\alpha}t^2)\right), \tag{3}$$

where $\alpha = \alpha_1 + i\alpha_2$ is related to the displacement of the vortex core from the disk center, $t_c = \left(1 - \sqrt{1 - 4\alpha\bar{\alpha}}\right)/(2\bar{\alpha})$. Equations (1-3) then represent the magnetization for the vortex confined to the circular disk with the vortex center at $t_c$.

Next consider the mapping of the circular disk in the $t$ plane onto the ellipse in the complex $z = x + iy$ coordinate system also with no edge magnetostatic charge. This is the Riemann-Hilbert problem [20] for the transformation $z = M(t)$ and the function in Eq. (2) giving the magnetization in the ellipse is

$$f(z) = M'(t)f(t), \tag{4}$$

where the prime indicates a derivative. The mapping of the circle to the ellipse was first found by Schwarz [21], and it was later derived in a simpler way [22]. Here it is assumed that the ellipse is in the complex $z$ plane with foci at $\pm 1$ and semi-axes satisfy the condition $a^2 - b^2 = 1$. For the ellipse in the complex $z$ plane, the Schwarz formula is expressed in terms of the Jacobi elliptic function

$$t = k^{1/2} sn\left(\frac{2k'}{\pi}\sin^{-1} z\right). \tag{5}$$

Here the modulus $k$ can be expressed in terms of the theta function, but a more useful form for numerical calculations is

$$k = \left(\frac{2q^{1/4} + 2q^{9/4} + 2q^{25/4} + \cdots}{1 + 2q + 2q^4 + 2q^9 + \cdots}\right)^2, \tag{6}$$

and the parameter is $q = 1/(a+b)^4$. The modulus and complementary modulus, $k' = \sqrt{1-k^2}$ are defined according to Mathematica for numerical calculations.

Next let us obtain the magnetization in the ellipse. First use Eq. (5) explicitly giving $t(z)$ as a function of $z$, which is used in Eq. (3) to obtain $f(t(z))$ also as a function of $z$. Now we use the inverse of Eq. (5)

$$z = \sin\left[\frac{\pi}{2k'}sn^{-1}(t/\sqrt{k})\right] = M(t), \qquad (7)$$

which can now be used to simply evaluate the derivative, $M'(t)$ in Eq. (4). It is remarked that the inverse of the elliptic function is the elliptic integral of the first kind. To finally find the magnetization in the ellipse, Eq. (5) is substituted into this derivative and Eq. (4) is used to get $f(z)$, then used in Eqs. (1 and 2) giving the structure of the magnetic vortex in the ellipse. Following the above steps, the form of the magnetization is obtained in a straightforward manner, but it is too long to include here. This structure is illustrated in Fig. 1 showing the vortex with its core slightly displaced from the ellipse center and the magnetization is parallel to the ellipse edge.

### III. Magnetostatic Energy

For the nanoparticle the exchange energy is minimized owing to Eq. (1) and the vortex dynamic properties are mainly determined [1,3] by the magnetostatic contribution to the energy. For the case of the circular disk the MS energy has a quadratic form, $W_{MS} = \kappa X^2/2$. In the following the MS energy of the ellipse is obtained as a quadratic function of $X$, but owing to the lower symmetry the energy is nonzero when $X = 0$; moreover, it will also be necessary to consider displacements along the $y$ axis.

Let us consider a thin nanoparticle with the MS energy obtained from the magnetostatic field, $\vec{h}$

$$W_{MS} = -\frac{\mu_0 M_s^2}{2} \int \vec{m} \cdot \vec{h} \, d^3r, \tag{8}$$

where the integration is over the ellipse volume. The components of the reduced magnetic field are calculated from the MS potential, $\vec{h} = \partial \Phi_{MS}/\partial \vec{r}$, with

$$\Phi = \int \frac{\nabla \cdot \vec{m}(\vec{r}')}{|\vec{r}-\vec{r}'|} d^3r'. \tag{9}$$

The integrals over $z$ and $z'$ in Eqs. (8 and 9) can be done for the case of the thin disk ($L \ll b$) when the magnetization is assumed to be independent of $z$ to obtain a factor of $L^2$ in Eq. (8). In this case the reduced components, $h_{x,y}$ of the magnetostatic field are

$$h_x = \int \frac{(x-x')\nabla \cdot \vec{m}(x',y')}{[(x-x')^2+(y-y')^2]^{3/2}} dx'dy' \tag{10a}$$

$$h_y = \int \frac{(y-y')\nabla \cdot \vec{m}(x',y')}{[(x-x')^2+(y-y')^2]^{3/2}} dx'dy', \tag{10b}$$

where the integrations is over the ellipse area.

Using Eqs. (10a,b) the MS energy is calculated as a four-dimensional numerical integral over the ellipse area using the normalized magnetization of the previous section as well as the above reduced magnetic field in Eq. (8). For the special case of the circular disk, the higher symmetry reduces the integration to one or more dimensions depending on the approximations used. The MS energy is expressed as a quadratic function of $X$, $W_{MS} = \mu_0 M_s^2 L^2 \kappa X^2/2$, where $\kappa = 8\pi^2 C/R$ and $C \approx 0.15$ calculated [1] by numerical integration. Because of the lower symmetry of the ellipse a four-dimensional integral must be done, and for small vortex core displacements from the ellipse center the integral in Eq. (8) has the form,

$$w(X,Y) = \int (m_x h_x + m_y h_y) dxdy. \tag{11}$$

Here $X$ is the displacement of the vortex center along the $x$ axis and $Y$ is the displacement along the $y$ axis. Of course the $X$ and $Y$ dependence is contained in the volume magnetostatic charge, $\nabla \cdot \vec{m}$ which increases as the vortex core moved off center and Eq. (11) is a reduced MS energy having units of length. For the displaced vortex bear in mind that $t_c$ defined immediately following Eq. (3) is the position of the vortex core in a circular disk. This displacement must be transformed to the ellipse to give $z_c = X + iY$ using the transformation given by Eq. (7) where $X$ and $Y$ are functions of $\alpha_1$ and $\alpha_2$. It will only be necessary to consider the cases $\alpha_2 = 0$ to calculate the quadratic $X$ terms or $\alpha_1 = 0$ to calculate the quadratic $Y$ terms since there are no cross-terms owing to symmetry. The numerical integrals defined by Eq. (11) are quadratic functions of $X$ and $Y$ expressed as

$$W_{MS} = \frac{\mu_0 M_s^2 L^2}{2}(w_0 + w_1 X^2 + w_2 Y^2), \qquad (12)$$

with the coefficients depending only on the ellipse parameters and both $w_1$ and $w_2$ have units of 1/length. Because of the lower symmetry the first term is independent of vortex displacement, and there are different coefficients corresponding to the $X$ and $Y$ displacements of the vortex core. To obtain the coefficients in Eq. (12) the numerical integration is done for various small values of $X$ or $Y$ and a straight line plot of these integral values versus $X^2$ or $Y^2$ is used to estimate $w_1$, and $w_2$. These coefficients are calculated as a function of the ellipticity, $b/a$.

First the energy of the centered vortex with $X = Y = 0$ is calculated. Since energy calculations are subject to the condition $a^2 - b^2 = 1$, the volume of the ellipse as well as the MS energy are increasing functions of $b/a$, therefore, the energy density, $w_{MS} = w_0/\pi a b L$ is calculated to eliminate the volume dependence. This term is not important for the dynamic properties, but it is interesting to see the dependence of the energy density on the ellipticity as illustrated in Fig. 2.

Now let us determine the displacement $(X, Y)$ dependence of the magnetostatic energy. In the following, the integral evaluated in Eq. (11) with $Y = 0$ is labeled by $w_X$ and the integral with $X = 0$ is labeled by $w_Y$. Typical plots of these four-dimensional energy integrals versus vortex displacement are shown in Fig. 3 for $b/a = 0.816$ to get $w_1 = 7.06$ and $w_2 = 10.43$ from the slopes of these graphs. These parameters have been determined numerically for values of $b/a$ from 0.225 to 0.90 with the results shown in Fig. 4. It has been shown through dimensional analysis [10] and more recently through micromagnetic calculations [18] that the coefficients are related by

$$w_1 = w_2 \frac{b^2}{a^2}. \qquad (13)$$

This is also seen in Fig. 4 illustrating the calculated values of $w_1$ and $w_2 \frac{b^2}{a^2}$. Therefore, it is only necessary to find the $a$ and $b$ dependence of one parameter to specify a general form of the MS energy. The main goal here is a determination of the $a$ and $b$ dependence of the parameter, $w_1$. To accomplish this let us look for a function of $w_1$ including both $a$ and $b$ that will appear to be linear. Since $w_{1,2}$ have units of 1/length we plot $w_1$ with a length factor, and it is noticed that a plot of $w_1\sqrt{ab}$ versus $b/a$ gives the desired linear form as shown in Fig. 5. In both Figs. (4, 5) the value of $w_1$ is expected to go to a zero limit as $b/a \to 0$, but the vortex will no longer be a metastable state at some point as the ellipse is stretched into a needle structure; nevertheless, this limit is also included on the graphs. Notice here the linear dependence for $b/a > 0.3$ with nonlinearity for smaller values of the ellipticity. From this linear dependence, finally the parameter dependence of $w_1$ is given by

$$w_1(a,b) = \frac{1}{\sqrt{ab}}\left(-2.60 + 16.66\frac{b}{a}\right). \tag{14}$$

The main result of this work is contained in Eq. (14) from which the MS energy can be obtained for any value of $b/a$ in the range from about 0.3 extrapolated to 1. Finally it is pointed out that these numerical values of the MS energy in the previous figures were calculated for vortex displacements scaled relative to $c = 1$, but the scale does not have an effect on linear dynamics as in the next section.

## IV. Vortex Dynamics

For the vortex in a disk the largest amplitude mode corresponds to gyrotropic precession of the vortex core about the disk center typically described by the Thiele equation,

$$\vec{G} \times \vec{V} = \nabla W_{MS}, \tag{15}$$

where $\vec{G} = \frac{2\pi\mu_0 M_s L}{\gamma}\hat{z}$ is the gyroconstant and $\vec{V} = (\dot{X},\dot{Y})$ is the velocity of the vortex center with the dot indicating a time derivative. It is also stressed that this form for $\vec{G}$ is calculated by an integral over the thin nanostructure with the assumption that the vortex core is represented by a δ-function. The gyroconstant in Eq. (15) is a very good approximation when the vortex core is very small compared to the lateral dimensions of the nanoparticle. The gyrotropic frequency for the ellipse [18] has been shown to be $\omega = \sqrt{w_1 w_2}/G$ ; moreover, with the $a$ and $b$ dependence of both $w_1$ and $w_2$ known from Eqs. (13, 14), and also using $\dot{X} = \omega X, \dot{Y} = \omega Y$ the expression for the gyrotropic frequency of the ellipse vortex can be obtained. Let us finally use Eqs. (13, 14), as well as the expression for $G$ to obtain the general dependence of the gyrotropic frequency on the ellipse axes,

$$\omega_{GT} = \frac{\omega_M a}{8\pi^2 b} \frac{L}{\sqrt{ab}} \left(-2.60 + 16.66 \frac{b}{a}\right), \tag{16}$$

where $\omega_M = 4\pi M_s \gamma$ is a characteristic frequency ( about 15.1 GHz for permalloy [18]) and $\gamma$ is the gyromagnetic ratio. Finally the expression for the gyrotropic frequency for the circular disk in the limit where $a = b = R$ is found by extrapolation of Fig. 5 to $b/a = 1$ to get 14.1 for the value in parentheses, then the frequency has the form $\omega = \omega_M C L/R$ where $C \approx 0.18$ similar to previous [1] estimates. Previous experimental work [6, 9-11] involving measurements of vortex dynamics in magnetic ellipses had ellipticities within the range applicable to Eq. (16) with $b/a \sim 0.5$. The experimental measurements [6] of the gyrotropic frequency in the thick disk driven by spin polarized current for the disk with parameters $2a = 160$ nm, $2b = 75$ nm, and $L = 60$ nm is 1.2 GHz. Compare this with the calculated frequency using Eq. (16) which is about 2.4 GHz, which is over-estimated because this is definitely not a thin disk. The frequency calculation here is applicable to the data of Refs. [10, 11], which is a thin disk with $2a = 1000$ nm , $2b = 500$ nm and $l = 40$ nm. For this ellipse the experimental frequency was determined to be 0.116 GHz and the frequency calculated from Eq. (16) is 0.126 GHz. The frequency calculated in Ref. [10] using micromagnetic modeling was slightly higher at 0.134 GHz.

## V. Conclusion

The exchange and MS energies in the thin elliptical nanostructure are minimized using complex variable methods to obtain an expression for the MS energy valid in general for a range of ellipticity from about 0.3 to 1.0. The obtained expression for the gyrotropic frequency gives results that are in agreement with the limited experimental data available for the thin elliptic nanostructure. Other experimental publications have reported the observation of two vortices [9] in a single ellipse, and our current work is not applicable to this multi-vortex case. However, Eq.

(3) can be replaced by a two-vortex ansatz and the remaining calculations will be similar, but with a more complicated magnetic structure. In principle this technique can be extended to multiple vortices and antivortices such as two vortices interacting through an intermediate antivortex. In this case the MS energy will include vortex-vortex as well as vortex-antivortex interactions which will be the subject of future work.

## VI. References

[1] K. Yu. Guslienko, B. A. Ivanov, V. Novosad, Y. Otani, H. Shima, and K. Fukamichi, J. Appl. Phys. **91**. 8037 (2002).

[2] J. P. Park, P. Eames, D. M. Engebretson, J. Berezovsky and P. A. Crowell, Phys. Rev. B**67**,020403 (2003).

[3] K. Yu. Guslienko, X. F. Han, D. J. Keavney, R. Divan and S. D. Bader, Phys. Rev. Lett.**96**, 067205 (2006).

[4] M. Buess, M. Höllinger, T. Haug, K. Perzlmaier, U. Krey, D. Pescia, M. R. Scheinfein, D. Weiss, and C. H.Back, Phys. Rev. Lett.**93**, 077207 (2004).

[5] S. Kasai, Y. Nakatani, K. Kobayashi, H. Kohno, and T. Ono, Phys. Rev. Lett. **97**, 107204 (2006).

[6] V. S. Pribiag, I. N. Krivototov, G.. D. Fuchs, P. M. Braganca, O. Ozatay, J. C. Sankey, D. C. Ralph, and R. A. Buhrman, Nature Phys. **3**, 498 (2007).

[7] A. D. Belanovsky, N. Locatelli, P. N. Skirdkov, R. Abreu Araujo, F. Grollier, K. A. Zvezdin, V. Cros, and A. K. Zvezdin, Phys. Rev. B **85**, 100409 (2012).

[8] A. D. Belanovsky, N. Locatelli, P. N. Skirdkov, F. Abreu Araujo, K. A. Zvezdin, J. Grollier, V. Cros, and A. K. Zvezdin, Appl. Phys. Lett. **103**, 122405 (2013).

[9] K. S. Buchanan, P. E. Roy, M. Grimsditch, F. Y. Fradin, K. Yu. Guslienko, S. D. Bader, and V. Novosad, Nature Phys. **1**, 172 (2005).

[10] K. S. Buchanan, P. E. Roy, F. Y. Fradin, K. Yu. Guslienko, M. Grimsditch, S. D. Bader, and V. Novosad, J. Appl. Phys. **99**, 08C707 (2006).

[11] K. S. Buchanan, P. E. Roy, M. Grimsditch, F. Y. Fradin, K. Yu. Guslienko, S. D. Bader, and V. Novosad, Phys. Rev. B **74**, 064404 (2006).

[12] A. A. Thiele, Phys. Rev. Lett. **30**, 230 (1973).

[13] D. L. Huber, Phys. Rev. B **26**, 3758 (1982).

[14] K. Yu. Guslienko, W. Scholz, R. W. Chantrell, and V. Novosad, Phys. Rev. B **71**, 144407 (2005).

[15] K-S Lee, and S-K Kim, Appl Phys. Lett. **91**, 132511 (2007).

[16] K. Kim and S-B Choe, Journal of Magnetics **12**,113 (2007).

[17] R. Antos, Y. Otani, and J. Shibata, J. Phys. Soc. Japan **77**, 031004 (2008).

[18] G. M. Wysin, Low Temp. Phys.**41**, 1009 (2015).

[19] K. L. Metlov, Phys. Rev. B **88**, 014427 (2013).

[20] K. L. Metlov, Phys. Rev. Lett. **105**, 107201 (2010).

[21] H. A. Schwatz, J. für die reine und angew. Mathematik**70**, 105 (1869).

[22] G. Szego, The Am. Math. Monthly **57**, 474 (1950).

[23] B. A. Ivanov and C. E. Zaspel, J. Appl. Phys. **95**, 7444 (2004).

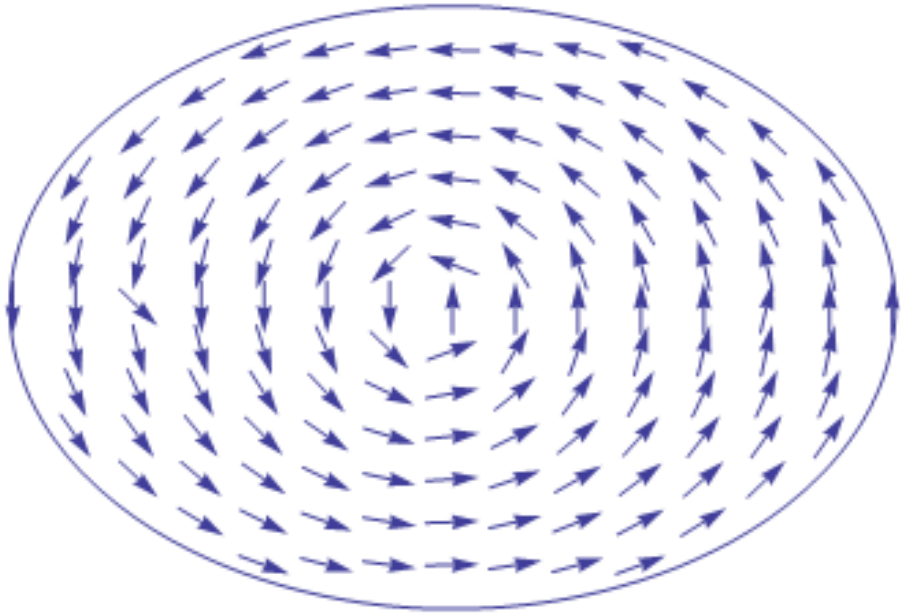

Fig. 1. Ellipse vortex inside of the $a = 1.41$, $b = .99$ ellipse having an ellipticity of 0.707. The vortex is off center along the $x$ axis with $a_1 = 0.05$ corresponding to a displacement of $X = 0.057$.

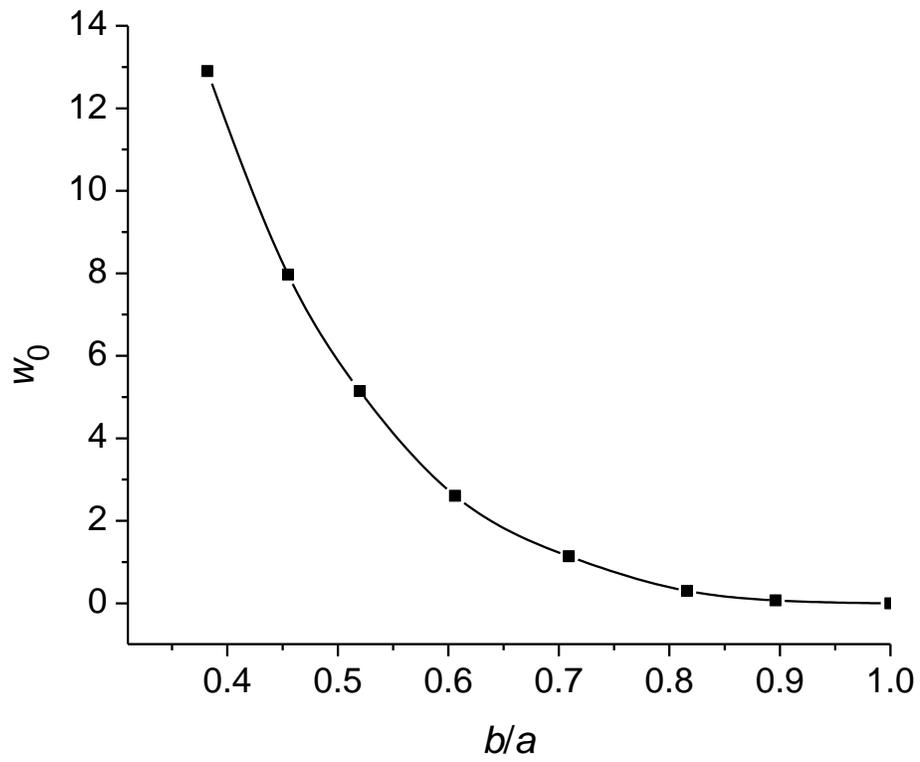

Fig. 2. Reduced energy of the centered ($X = Y = 0$) ellipse represented by $w_0$ in Eq. (12). The squares represent numerically calculated points and the curve is a guide to the eye.

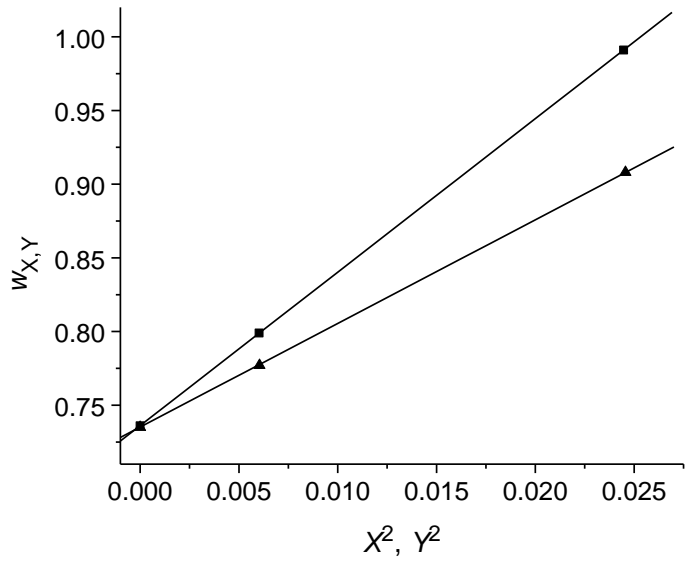

Fig. 3. Coefficients of the displacement-dependent terms of the reduced energy versus $X^2$ or $Y^2$. The energy integral $w_X$ given by the lower line plotted as a function of $X^2$ and $w_Y$ is given by the upper line plotted as a function of $Y^2$.

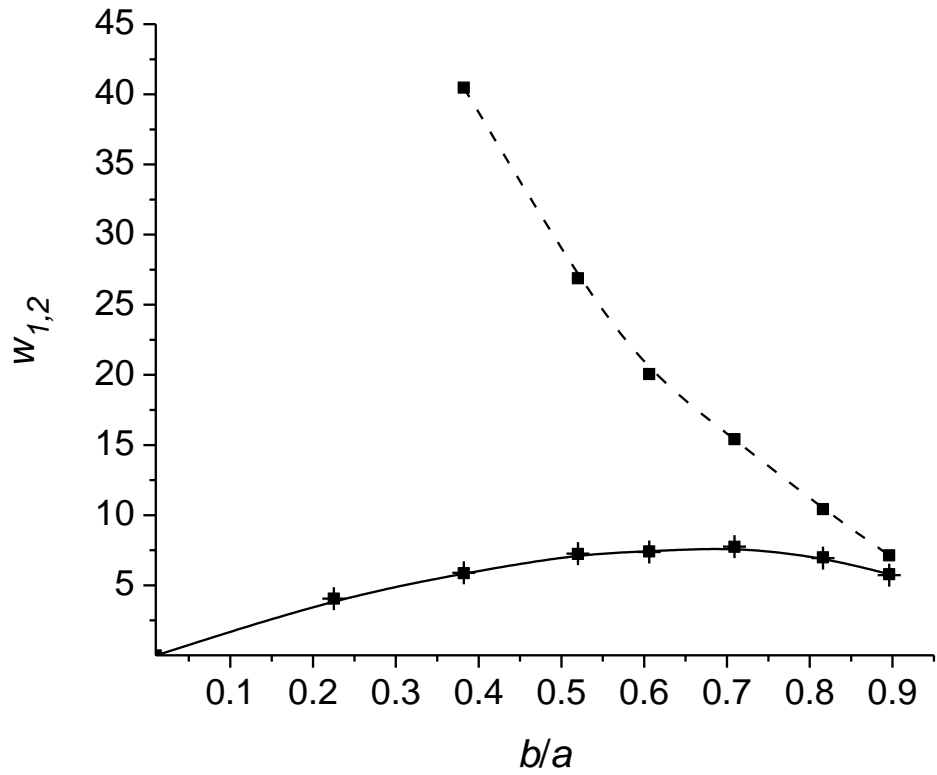

Fig. 4. Coefficients of $X^2$ ($w_1$, solid curve) and $Y^2$ ($w_2$, dashed curve) versus the ellipticity. The squares are the numerically calculated points and the crosses represent $w_2 b^2/a^2$.

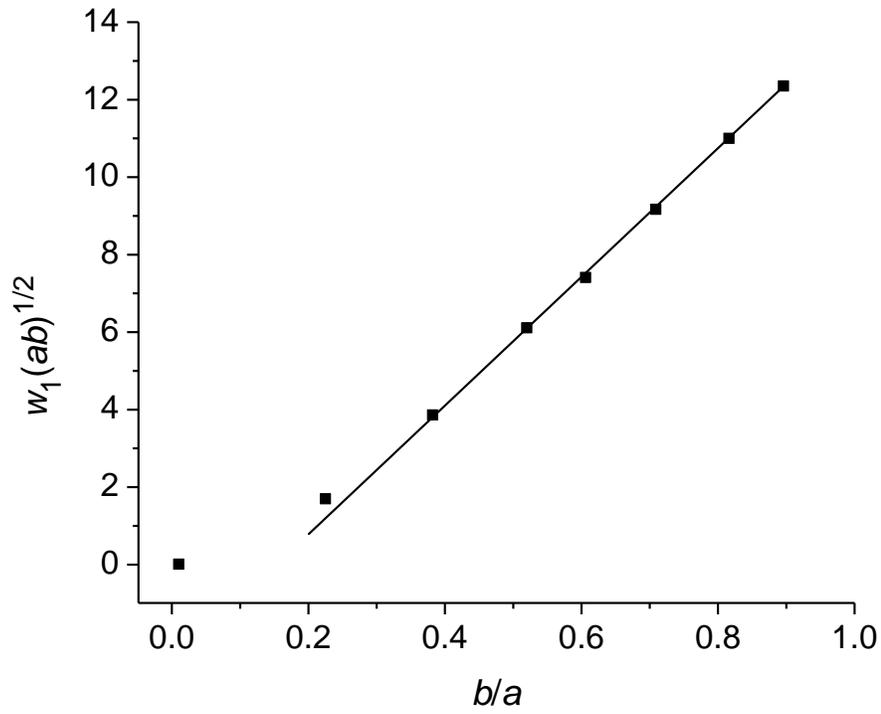

Fig. 5. Plot of $w_1\sqrt{ab}$ versus ellipticity. Squares represent calculated points and the straight line shows the region where the dependence is linear.